\documentclass[fleqn,usenatbib]{mnras}
\usepackage{newtxtext,newtxmath}
\usepackage[T1]{fontenc}
\usepackage{ae,aecompl}

\usepackage{graphicx}	

\newcommand{\Mwd}{\mbox{$\mathrm{M}_\mathrm{wd}$}}
\newcommand{\Msun}{\mbox{$\mathrm{M}_{\odot}$}}
\newcommand{\Rsun}{\mbox{$\mathrm{R}_{\odot}$}}
\newcommand{\Teff}{\mbox{$T_{\mathrm{eff}}$}}
\newcommand{\logg}{\mbox{$\log g$}}
\newcommand{\Ha}{\mbox{$\mathrm{H}\alpha$}}
\newcommand{\Hb}{\mbox{$\mathrm{H}\beta$}}
\newcommand{\Ion}[2]{#1\,{\sc #2}}

\usepackage[usenames,dvipsnames]{xcolor}
\usepackage{soul}

\title[Single magnetic white dwarfs with Balmer emission lines]{Single magnetic white dwarfs with Balmer emission lines: \\
A small class with consistent physical characteristics as possible signposts for  close-in planetary companions}

\author[G\"ansicke et al.]{Boris T. G\"ansicke$^{1}$\thanks{E-mail: Boris.Gaensicke@gmail.com},
Pablo Rodr\'iguez-Gil$^{2,3}$,
Nicola P. Gentile Fusillo$^{4}$, \newauthor
Keith Inight$^{1}$,
Matthias R. Schreiber$^{5,6}$,
Anna F. Pala$^{4}$,
Pier-Emmanuel Tremblay$^{1}$\\
$^{1}$Department of Physics, University of Warwick, Coventry, CV4 7AL, UK\\
$^{2}$Instituto de Astrof\'isica de Canarias, 38205 La Laguna, Tenerife, Spain\\
$^{3}$Departamento de Astrof\'isica, Universidad de La Laguna, 38206 La Laguna, Tenerife, Spain\\
$^{4}$European Southern Observatory, Karl Schwarzschild Stra{\ss}e 2, Garching, 85748, Germany\\
$^5$Departamento de F{\'i}sica, Universidad T{\'e}cnica Federico Santa Mar\'ia, Av. Espa\~{n}a 1680, Valpara{\'i}so, Chile \\
$^6$Millennium Nucleus for Planet formation, NPF,  Valpara{\'i}so, Chile 
}
\date{Accepted XXX. Received YYY; in original form ZZZ}
\pubyear{2019}

\begin{document}
\label{firstpage}
\pagerange{\pageref{firstpage}--\pageref{lastpage}}

\maketitle

\begin{abstract}
We report the identification of SDSS\,J121929.45+471522.8 as the third apparently isolated magnetic ($B\simeq18.5\pm1.0$\,MG) white dwarf exhibiting Zeeman-split Balmer emission lines. The star shows coherent variability at optical wavelengths with an amplitude of $\simeq0.03$\,mag and a period of 15.26\,h, which we interpret as the spin period of the white dwarf. Modelling the spectral energy distribution and \textit{Gaia} parallax, we derive a white dwarf temperature of $7500\pm148$\,K, a mass of  $0.649\pm0.022$\,\Msun, and a cooling age of $1.5\pm0.1$\,Gyr, as well as an upper limit on the temperature of a sub-stellar or giant planet companion of $\simeq250$\,K. The physical properties of this white dwarf match very closely those of the other two magnetic white dwarfs showing Balmer emission lines: GD356 and SDSS\,J125230.93$-$023417.7. We argue that, considering the growing evidence for planets and planetesimals on close orbits around white dwarfs, the unipolar inductor model provides a plausible scenario to explain the characteristics of this small class of stars. The tight clustering of the three stars in cooling age suggests a common mechanism switching the unipolar inductor on and off. Whereas Lorentz drift naturally limits the lifetime of the inductor phase, the relatively late onset of the line emission along the white dwarf cooling sequence remains unexplained.
\end{abstract}



\section{Introduction}

White dwarfs are the remnants of stars born with initial masses $\lesssim8-10$\,\Msun\ \citep[e.g.][]{smarttetal09-1, cummingsetal19-1}. Whereas the possibility that some of these stellar remnants possess strong magnetic fields was explored already by \citet{blackett47-1}, the observational confirmation occurred only much later \citep{kempetal70-1}. It is now firmly established that a small fraction, $2-10$~per cent of single white dwarfs exhibit  magnetic fields of $B\gtrsim1$\,MG  \citep{hollandsetal15-1, ferrarioetal15-1, kawka20-1}, and there is evidence that weaker fields are equally or possibly even more common \citep{landstreet+bagnulo19-1,landstreet+bagnulo19-2,bagnulo+landstreet19-1}. The origin of magnetic fields in white dwarfs is still debated, with working hypotheses including fossil fields \citep{angeletal81-1, braithwaite+spruit04-1}, binary interactions either in the form of a common envelope (\citealt{toutetal08-1}, but see \citealt{bellonietal20-1} for a discussion of a number of serious problems in that scenario) or mergers \citep{garcia-berroetal12-1}, or processes internal to the white dwarf \citep{isernetal17-1}.

In the vast majority of magnetic white dwarfs, the presence of the field is established via the detection of Zeeman-split absorption lines of the atmospheric constituents: hydrogen \citep{angeletal74-2}, helium \citep{jordanetal98-2}, carbon \citep{schmidtetal99-1}, and other metals \citep{kawka+vennes11-1}. 

One exception to this rule has been the maverick white dwarf GD356, exhibiting  Balmer \textit{emission} lines Zeeman-split in a field of $B\simeq11$\,MG \citep{greenstein+mccarthy85-1}. No binary companion has been detected \citep{ferrarioetal97-2}, and speculative scenarios explaining the origin of the emission lines included convective activity \citep{greenstein+mccarthy85-1} or accretion from the interstellar medium \citep{ferrarioetal97-2}. However, \citet{weisskopfetal07-1} argued against a hot corona of any kind based on the non-detection of either X-rays or cyclotron radiation. Time-series photometry revealed quasi-sinusoidal variability with an amplitude of 0.2~per cent and a period of $\simeq115$\,min, interpreted as the spin period of the white dwarf \citep{brinkworthetal04-1}, which is moderately rapid for a single white dwarf \citep{hermesetal17-1}.

Left with none of the conventional models providing a satisfactory explanation for the Balmer emission lines, \citet{lietal98-1} suggested that a conductive planet, or planet core in a close orbit around GD356 would result in the generation of electric currents that could heat the regions near the magnetic poles of the white dwarf~--~akin to the Jupiter-Io configuration \citep{goldreich+lynden-bell69-1}. \cite{wickramasingheetal10-1} revisited the unipolar inductor model, arguing that such a planet was unlikely to have survived the giant branch evolution of the progenitor of GD356. Instead, these authors proposed that GD356 is the product of a double-white dwarf merger and the putative planet formed from the metal-rich debris disc left over by this merger \citep{garcia-berroetal07-1}~--~a rare event, explaining the (at the time) unmatched properties of GD356. 

In the light of the rapidly growing evidence for planetesimals and planets around white dwarfs \citep{becklinetal05-1, gaensickeetal06-3, farihietal09-1, vanderburgetal15-1, manseretal19-1, gaensickeetal19-1}, \cite{veras+wolszczan19-1} studied the survivability of conductive planetary cores, and found that a significant parameter space of white dwarf plus planet configurations exists, lending support to the unipolar inductor model.  

GD356 remained a fascinating but lonely system for 35 years, until \citet{redingetal20-1} announced the discovery of  SDSS\,J125230.93$-$023417.7 (SDSS\,J1252$-$0234), a second magnetic ($B\simeq5$\,MG), single white dwarf exhibiting Zeeman-split Balmer emission lines with an exceptionally short rotation period of 317\,s~--~making it the fastest spinning white dwarf. 

We report the identification of a third magnetic ($B\simeq18.5$\,MG), single white dwarf with Zeeman-split Balmer emission lines, SDSS\,J121929.45+471522.8. We also discuss the physical properties and possible nature of this emerging new class of white dwarfs and their possible link to close planetary companions.

\section{Observations}

\subsection{Spectroscopy}
\begin{figure}
\centering{\includegraphics[width=\columnwidth]{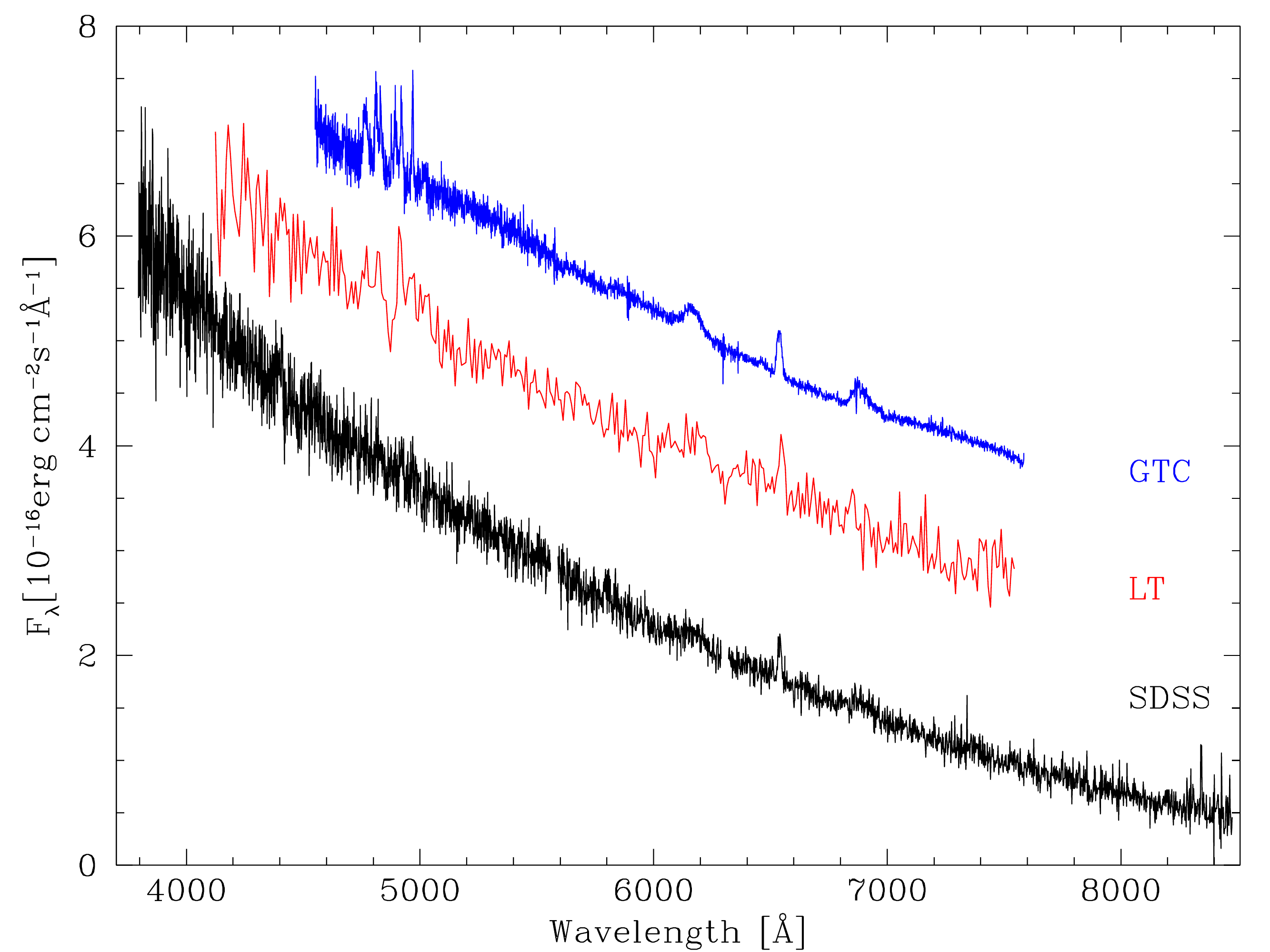}}
\caption{Spectroscopy of SDSS\,J1219+4715. The SDSS spectrum (April 2004) revealed weak \Ha\ emission, which was confirmed with the LT in April 2020. Higher resolution spectra obtained with the GTC between April and June 2020 show the Zeeman-split emission lines of \Ha\ and \Hb.}
\label{f:spectra}
\end{figure}

The first spectrum of SDSS\,J121929.45+471522.8 (SDSS\,J1219+4715 henceforth) was obtained on 2004 April 21 with the SDSS spectrograph on the 2.5-m SDSS telescope (Fig.\,\ref{f:spectra}; \citealt{yorketal00-1, adelman-mccarthyetal06-1}), using an exposure time of 2520\,s. The SDSS spectrum covers the wavelength range $3800-9200$\,\AA\ at a spectral resolution of $R=\lambda/\delta\lambda\simeq1800$.  \citet{szkodyetal06-1} noticed a weak \Ha\ emission line in this spectrum and classified SDSS\,J1219+4715 as a candidate cataclysmic variable (CV), i.e. a short-period interacting binary containing a white dwarf and a low-mass star. The authors argued more specifically that the optical spectrum is likely dominated by the hot accretion disc of a CV with a relatively high mass transfer rate. This hypothesis was ruled out by \citet{palaetal20-1} based on the \textit{Gaia} parallax \citep{gaiaetal18-1}, putting the system firmly at a distance of $d=69.9\pm0.6$\,pc \citep{bailer-jonesetal18-1}. \citet{palaetal20-1} noted that the wavelength of the emission feature is $\simeq6540$\,\AA, noticeably blue-ward of \Ha, and tentatively explained it by contamination from the nearby large Seyfert~2 galaxy M\,106. 

To clarify the nature of SDSS\,J1219+4715, we obtained one spectrum on 2020 April 21 using the SPectrograph for the Rapid Acquisition of Transients (SPRAT, \citealt{piasciketal14-1}) on the robotic Liverpool Telescope (LT, \citealt{steeleetal04-1}). These data cover the wavelength range $\simeq4000-8000$\,\AA\ with a dispersion of 4.6\,\AA\ per pixel, resulting in $R\simeq340$ at 6000\,\AA. The exposure time was 1800\,s. The SPRAT observations were processed by an automated pipeline which corrects for bias, dark and flatfield effects, performs the sky subtraction and extracts the spectrum, and derives the wavelength and flux calibration. Cosmic rays were manually cleaned from the reduced spectrum. Despite the low resolution, the LT spectrum (Fig.\,\ref{f:spectra}) confirmed the emission line near \Ha\ noticed previously in the SDSS spectrum \citep{szkodyetal06-1}. The wavelength of the emission feature is $\simeq6540$\,\AA, also blue-ward of \Ha, and within uncertainties identical to that in the SDSS spectrum. Interpreting this line as Doppler-shifted \Ha\ emission would imply a velocity of $\simeq1000$~km\,s$^{-1}$, extremely high for any kind of close white dwarf binary. Moreover, it would be rather unlikely that our LT spectrum sampled the same orbital phase as the SDSS spectrum obtained 16 years earlier. Closer inspection of the LT spectrum suggested that additional structure was present near \Hb. 

Intrigued by the features in the LT spectrum, we acquired higher resolution spectroscopy with the Optical System for Imaging and low-Intermediate-Resolution Integrated Spectroscopy \citep[OSIRIS;][]{sanchezetal12-1} on the 10.4-m Gran Telescopio Canarias (GTC). The observations were carried out as 600\,s exposures in the period April 24 to 2020 June 29. We obtained a total of two and four spectra using the R2500V and R2500R grisms, respectively, which provide $R\simeq2000$ with the 1-arcsec slit. The wavelengths covered were $4500-6000$\,\AA\ and $5575-7685$\,\AA\ for the R2500V and R2500R grisms, respectively. 

The GTC spectra were reduced using {\sc iraf}\footnote{{\sc iraf} is distributed by the National Optical Astronomy Observatories.}. After debiasing and flat-fielding, we performed cosmic-ray removal with the L.A.Cosmic package \citep{vandokkum01-1}. Optimal extraction of the spectral trace \citep{horne86-1} was subsequently done with the {\sc starlink}/{\sc pamela} reduction package \citep{marsh89-1}. We used spectra of HgAr+Ne+Xe arc lamps obtained at the beginning of the nights for wavelength calibration, which was performed with {\sc molly}\footnote{{\sc molly} is available at \url{http://deneb.astro.warwick.ac.uk/phsaap/software}}. 

\begin{figure*}
\centering{
\includegraphics[width=0.8\textwidth]{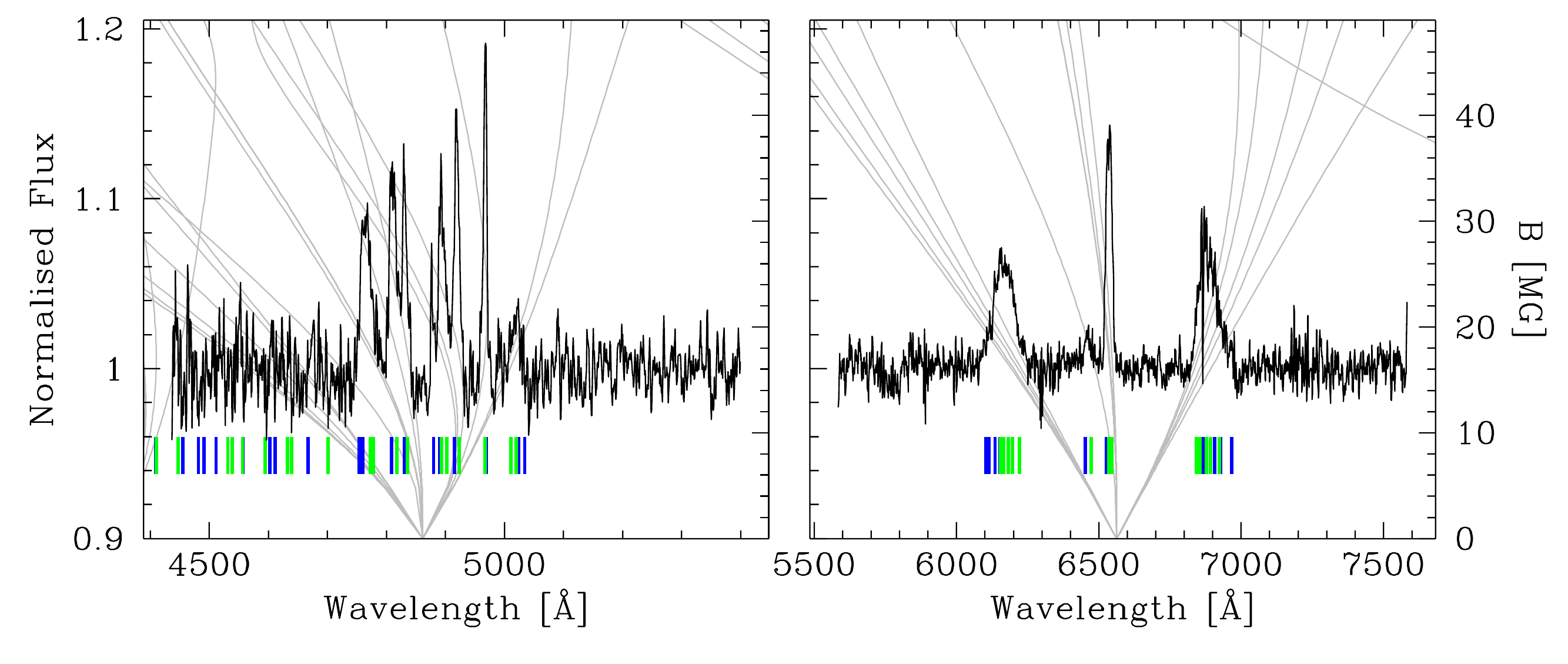}}
\caption{Normalised and averaged GTC spectra of SDSS\,J1219+4715, centred on \Hb\ (left) and \Ha\ (right). Shown in grey are the wavelengths of the individual Balmer line components as a function of field strength, which is given on the right-hand-side axis. The field strength in SDSS\,J1219+4715 is near the transition between linear and quadratic Zeeman splitting for \Ha, and in the quadratic regime for \Hb. The tick marks below the spectrum indicate the locations of the Balmer transitions in a field of 17.5\,MG (green) and 19.5\,MG (blue), illustrating that the spread in field strength within the region where the emission lines form is small, $B\simeq18.5\pm1.0$\,MG.}
\label{f:bfield}
\end{figure*}

The GTC  spectroscopy reveals that the emission detected previously in the SDSS and LT data is the $\pi$ component of Zeeman-split \Ha\ emission, flanked to the blue and red by the associated $\sigma^{-,+}$ components (Fig.\,\ref{f:spectra}, \ref{f:bfield}). Multiple sharp emission features are also detected near \Hb, unambiguously identifying SDSS\,J1219+4715 as a magnetic white dwarf.

\subsection{Photometry}
Given that many magnetic white dwarfs exhibit photometric variability \citep{hermesetal17-2}, we obtained sparse photometry with the robotic LT from 2020 May 19 to 2020 July 6. We used the Bessel $B$-band filter which provides high throughput over the range $\simeq3800-4900$\,\AA, covering the higher Balmer lines where photometric variability might be expected. We used the default detector binning of $2\times2$ and obtained groups of three 80-s exposures separated by at least two hours. We collected a total of 174 images over 38 individual nights. The LT data are provided in a reduced (bias-corrected and flat-fielded) format, and we extracted the photometry of SDSS\,J1219+4715 using the pipeline described in \citet{gaensickeetal04-1} relative to Gaia~DR2 1545014673992066048 ($G_\textrm{BP}=13.86$), a spectroscopically confirmed G-type star (LAMOST J121936.84+471554.8, \citealt{luoetal15-1,luoetal19-1}). The photometry revealed SDSS\,J1219+4715 to be variable with an amplitude of $\simeq 0.03$\,mag.

We complemented our LT observations with the $g$- and $r$-band photometry provided by Data Release~3 of the Zwicky Transient Facility (ZTF, \citealt{mascietal19-1}), adding 263 observations spanning 2018 April 9 throughout 2019 December 29. 

\section{Analysis}

\subsection{Magnetic field}
The GTC spectra of SDSS\,J1219+4715 clearly resolve three components of the \Ha\ emission line (Fig.\,\ref{f:bfield}, right panel), which can be identified as the $\sigma^-$, $\pi$, and $\sigma^+$ triplet of \Ha\ within the linear regime of the Zeeman effect, i.e. where the external field removes the energy degeneracy with respect to the magnetic quantum number $m_l$. However, the different widths of the three components imply that \Ha\ emission originates within an environment with a field strength $B$ where the Zeeman effect transitions into the quadratic regime, removing also the energy degeneracy with respect to the orbital angular momentum $l$, splitting \Ha\ into 15 transitions (of which two have degenerate energies, hence only 14 components can be observationally be detected, \citealt{henry+oconnell85-1}). The GTC spectra reveal a complex structure of multiple sharp emission features of \Hb\ (Fig.\,\ref{f:bfield}, left panel), which transitions into the quadratic Zeeman regime at lower fields compared to \Ha. We over-plot in Fig.\,\ref{f:bfield} the wavelengths of the individual \Ha\ and \Hb\ transitions \citep{friedrichetal96-1} as a function of $B$ and find that the \Ha\ emission arises in a field of $\simeq18.5$\,MG. Whereas as mentioned above, \Ha\ still largely appears as a linear Zeeman triplet, we do detect the $\mathrm{2p_0\rightarrow 3s_0}$ transition which is split off the other $\pi$ components.

The Zeeman components of \Ha\ and \Hb\ shift rapidly as a function of changing $B$-field, hence any substantial spread in the $B$-field across the region in which the emission lines arise would smear them out in wavelength. The sharpness of the \Ha\ $\pi$ components and of the \Hb\ emission lines implies a homogeneous field within the emitting region. We indicate in Fig.\,\ref{f:bfield} the locations of the individual Zeeman components for $B=17.5$\,MG (green ticks) and $B=19.5$\,MG (blue ticks), and conclude that the \Ha\ and \Hb\ emission lines arise within a field of $B\simeq18.5\pm1.0$\,MG. For a dipolar field structure on the white dwarf, the field near the magnetic equator is a factor two lower than the polar field \citep{achilleosetal92-2}, which implies that the emission lines form in a relatively small region(s), most likely near the magnetic pole(s). This is very similar to the situation in GD356, where \citet{greenstein+mccarthy85-1} measured $B=11\pm1.1$\,MG in the emitting region and \citet{ferrarioetal97-2} estimated that this region covers about ten per cent of the white dwarf surface. Similarly, \citet{redingetal20-1} derived $B=5.0\pm0.1$\,MG from the emission lines in SDSS\,J1252$-$0234, again implying a small spread in field strength across the region forming the emission lines. 

We note that whereas the individual GTC spectra do show some variation in the strength of the \Ha\ emission (Section\,\ref{s:variability}), we do not detect photospheric absorption lines from the white dwarf. At a field strength of $\simeq18.5$\,MG, the splitting of the three Zeeman components would by far exceed the pressure-broadening of the Balmer lines, in particular for a white dwarf as cool as SDSS\,J1215+4715, and hence the photospheric Balmer lines are usually easily visible (see e.g. fig.\,9 of \citealt{greenstein+mccarthy85-1} and fig.\,A12 of \citealt{tremblayetal20-1}). Most likely, the photospheric absorption  lines are filled in by the emission lines and possibly some small amount of continuum flux originating from the same region~--~which in SDSS\,J1215+4715 remains visible throughout the spin cycle. We note that in SDSS\,J1252$-$0234 the emitting region is self-eclipsed by the white dwarf for parts of the spin cycle, when the emission lines disappear and the photospheric Balmer absorption lines become visible \citep{redingetal20-1}.

\subsection{White dwarf parameters} 
\label{s:wdparam}
The standard technique to measure the atmospheric parameters, effective temperature (\Teff) and surface gravity (\logg), of hydrogen-atmosphere white dwarfs from modeling their Stark-broadened Balmer absorption lines \citep[e.g.][]{bergeronetal92-1, finleyetal97-1} is problematic in magnetic white dwarfs, as the simultaneous effect of electric and magnetic fields on the energy levels of hydrogen cannot yet be satisfactorily computed \citep{jordan92-1, friedrichetal94-1}. In the case of the three magnetic white dwarfs discussed here, a spectroscopic analysis of their atmospheric parameters is further complicated by the fact that the Balmer emission lines partially or fully fill in the photospheric absorption lines.

Nonetheless, at the moderate magnetic field strengths found at these white dwarfs, the overall spectral energy distribution of a star should not be significantly affected and, given an accurate distance measurement, \Teff\ and \logg\ can be reliably determined from broad-band photometry alone \citep[e.g.][]{koesteretal79-1} when also making use of the well-established mass-radius relation of white dwarfs \citep{paneietal00-2, tremblayetal17-1, parsonsetal17-1, joyceetal18-1,vedantetal20-1}. \citet{gentile-fusilloetal19-1} adopted this method to estimate atmospheric parameters for GD356, SDSS\,J1252$-$0234, and SDSS\,J1219+4715 using the \textit{Gaia} photometry and astrometry, and non-magnetic white dwarf model atmospheres. In the temperature range of these three stars, non-magnetic white dwarfs with hydrogen-rich atmospheres develop convection zones.  The effect of magnetic fields on convection has been discussed at length for solar-like stars \citep[e.g.][]{weiss66-1, chaplinetal11-1}. In the context of white dwarfs, it has been argued both on theoretical \citep{tremblayetal15-1} and observational grounds \citep{gentile-fusilloetal18-1} that the presence of fields in excess of $B\simeq50-100$\,kG suppresses convection, which results in an altered temperature structure within the atmosphere compared to weakly- or non-magnetic white dwarfs. For the fields considered here ($B\simeq5-18$\,MG, Table\,\ref{t:3mwd}), convection will be fully suppressed.

We have therefore re-derived the atmospheric parameters of the three stars fitting non-magnetic, purely radiative model spectra computed by enforcing a convective flux of zero when solving for the atmospheric stratification \citep{gentile-fusilloetal18-1}. We retrieved \textit{GALEX} (\textit{nuv}, \citealt{martinetal05-1}), SDSS (\textit{ugriz}, \citealt{albaretietal17-1}) and PanSTARRS (\textit{grizy}, \citealt{chambersetal16-1}) photometry for GD356, SDSS\,J1252$-$0234, and SDSS\,J1219+4715 and compared it with synthetic magnitudes calculated from our radiative models, scaled using \textit{Gaia} parallaxes and corrected for reddening using $E(B-V)$ values from the 3D STructuring by Inversion the Local Interstellar Medium (STILISM) reddening map \citep{lallementetal18-1}. We used a $\chi^2$ minimisation routine to find the best fitting models and determined the \Teff\ and \logg\ of the three white dwarfs.  As part of this procedure, we noted a marked discrepancy between the observed and predicted photometry in the SDSS-$u$ band. This filter covers the wavelength range  $3050-4000$\,\AA\ and is therefore very sensitive to flux changes in the Balmer jump, which are likely to arise from two separate effects. On the one hand, the presence of a magnetic field likely affects the profile of the Balmer jump in a way our models cannot reproduce. On the other hand, it is likely that the region responsible for the emission lines contributes also some small level of continuum flux associated with the bound-free opacity of hydrogen, $\kappa_\mathrm{bf}$. As $\kappa_\mathrm{bf}\propto\lambda^3$ up to the ionisation threshold of the considered energy level, this continuum emission is expected to  contribute in particular at wavelengths blue-wards of the Balmer jump. We therefore decided to exclude the SDSS-$u$ band photometry in our final fits. 

The atmospheric parameters are reported in Table\,\ref{t:3mwd}, along with the corresponding white dwarf masses and cooling ages, which we calculated using evolutionary models for thick hydrogen layers\footnote{\label{fn:bergeron} We use the 2020 version of the cooling models of \citet{holberg+bergeron06-1, kowalski+saumon06-1, fontaineetal01-1, tremblayetal11-2}, available at \href{http://www.astro.umontreal.ca/~bergeron/CoolingModels}{http://www.astro.umontreal.ca/$\sim$bergeron/CoolingModels}.}. We discuss the interpretation of the cooling ages in the context of the possible evolutionary history of these stars further in Sect.\,\ref{s:merger}. We note that the parameters derived here for SDSS\,J1252$-$0234 ($\Teff=7856\pm101$\,K, $\log g=7.98\pm0.06$) differ somewhat from those of \citet{redingetal20-1} ($\Teff=8237\pm206$\,K, $\log g=8.09\pm0.05$), who used convective atmosphere models, and did not make use of the \textit{GALEX} $nuv$ fluxes. 

We caution that there is one caveat to the white dwarf parameter determination: additional continuum flux associated with the observed emission lines will affect the fits, most likely resulting (via the scaling factor) in slightly overestimated radii, and hence underestimated masses. SDSS\,J1252$-$0234 displays the largest amplitude photometric variability among the three magnetic white dwarfs analysed here ($\simeq9$~per cent peak-to-peak in a BG40 blue broad bandpass) and is also the only system where the emission lines  disappear for a part of the white dwarf spin cycle, suggesting that the emitting region is self-eclipsed behind the white dwarf. Taking nine per cent as a conservative upper limit on the continuum flux contribution (some of the observed photometric variability may be associated with inhomogeneous emission across the surface of the magnetic white dwarfs), our white dwarf radii might be over-estimated by three per cent, corresponding to masses under-estimated by up to $\simeq0.028$\,\Msun. We note that this is the most extreme case, adopting the largest amplitude observed among the three stars and assuming that it is entirely related to additional emission in excess of the photospheric white dwarf flux. A quantitative assessment of this potential effect will be extremely difficult as it will require high-quality, accurately flux-calibrated, spin-phase resolved spectroscopy, and physically correct models for both the magnetic white dwarf photosphere and the emission region. 

\begin{figure}
\centering{\includegraphics[width=\columnwidth]{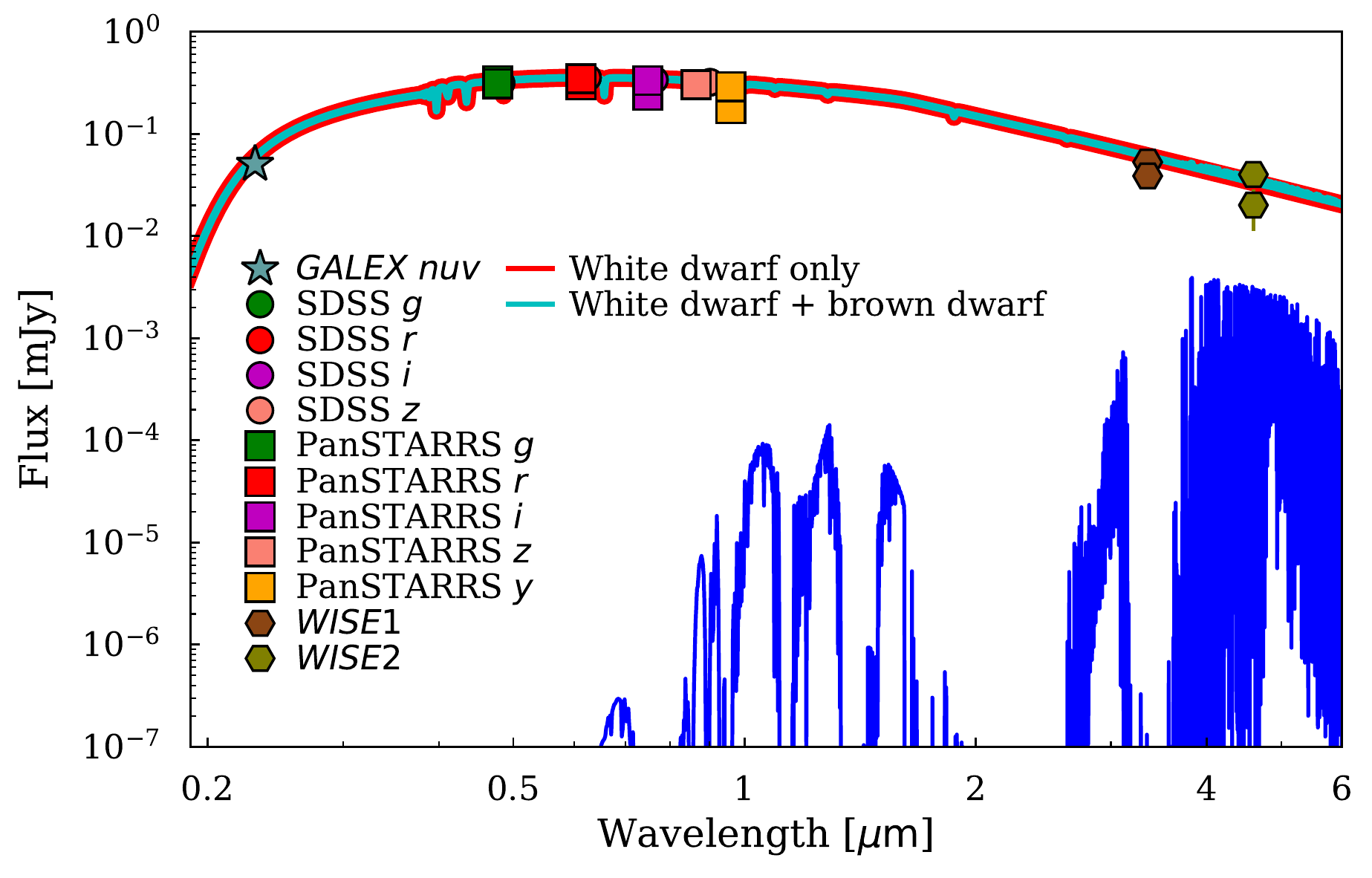}}
\caption{Spectral energy distribution of SDSS\,J1219+4715 showing the photometry obtained by the different surveys (coloured points). Shown in cyan is a composite model of the best–fit white dwarf ($T_\mathrm{eff,wd} = 7500\,$K, $\log g=8.09$, see Section~\ref{s:wdparam}) and a brown dwarf companion (blue) with effective temperature $T_\mathrm{eff,bd} \simeq 250\,$K, while the red curve shows the white dwarf only.}
\label{f:sed}
\end{figure}

\subsection{Limits on a possible companion}
\label{s:companion}
There is no evidence for a stellar companion at optical wavelengths. To place an upper limit on the presence of a substellar or planetary companion, we complemented the \textit{GALEX} (\textit{nuv}),  SDSS ($griz$), and Pan-STARRS ($grizy$) photometry used in Section\,\ref{s:wdparam} (where we exclude again the SDSS-$u$ band for the reasons discussed above) with the \textit{WISE} ($W1, W2$) bands \citep{wrightetal10-1}. The reconstructed spectral energy distribution of SDSS\,J1219+4715 was fitted using a model accounting for both the white dwarf flux and that of a possible low-mass stellar or substellar companion.

For the white dwarf, we used the best-fit radiative model spectrum ($T_\mathrm{eff,wd} = 7500\,$K, $\log g=8.09$, see Table\,\ref{s:clustered}). For the companion, we retrieved a grid of AMES-Cond 2000 models \citep{allardetal01-1,baraffeetal03-1} for brown dwarfs from the Theoretical Spectra Web Server\footnote{\href{Theoretical Spectra Web Server}{http://svo2.cab.inta-csic.es/theory/newov2/index.php?models=cond00}}. The grid covered the range $T_\mathrm{eff} = 200 - 4000\,$K in steps of $100\,$K for $Z = \mathrm{Z}_\odot$ and a surface gravity $\log g =4$. The latter was evaluated considering WD\,0806--661, a white dwarf  with a similar age to SDSS\,J1219+4715, which has a wide planet-mass companion, $M_\mathrm{P} = 7\,\mathrm{M_{J}}$ \citep{luhmanetal11-1}. We then assumed a typical radius for the given mass and age, $R = 0.1\,\Rsun$ (see e.g. fig.~3 from \citealt{burrowsetal11-1}).

We performed the spectral fit using the Markov chain Monte Carlo (MCMC) implementation for Python, \textsc{emcee} \citep{foreman-mackeyetal13-1}, constraining the white dwarf and the companion to be located at the same distance. The best-fit model is shown in Fig.~\ref{f:ha_stack} (cyan) and implies a brown dwarf companion (blue) with effective temperature $T \simeq 250\,$K. However, only accounting for the presence of the white dwarf (red) allows to adequately reproduce the observed flux level in the \textit{WISE} filters. Our result thus represents an upper limit on the effective temperature of a possible companion to SDSS\,J1219+4715.

\begin{figure*}
\includegraphics[width=\columnwidth]{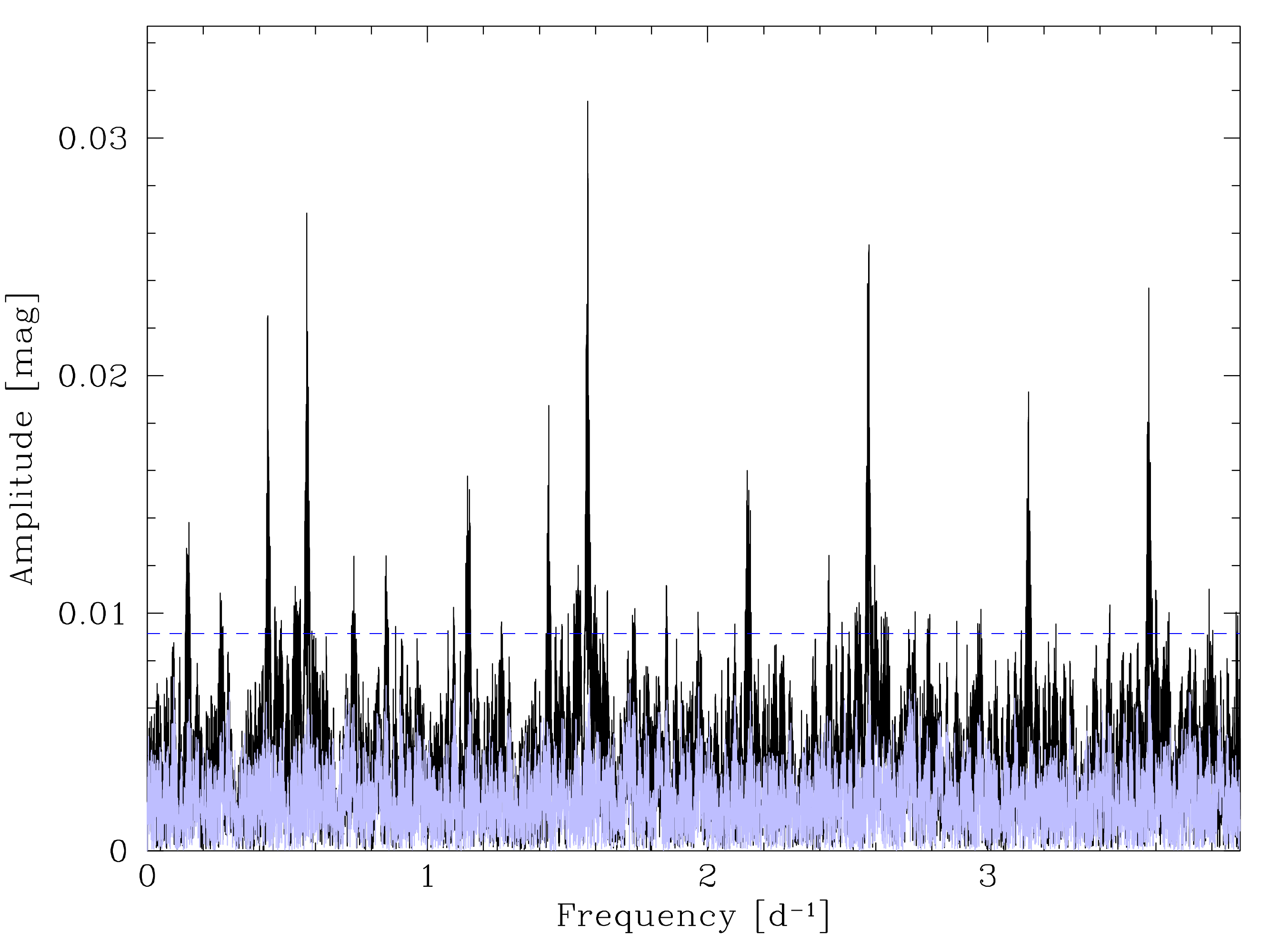}
\includegraphics[width=\columnwidth]{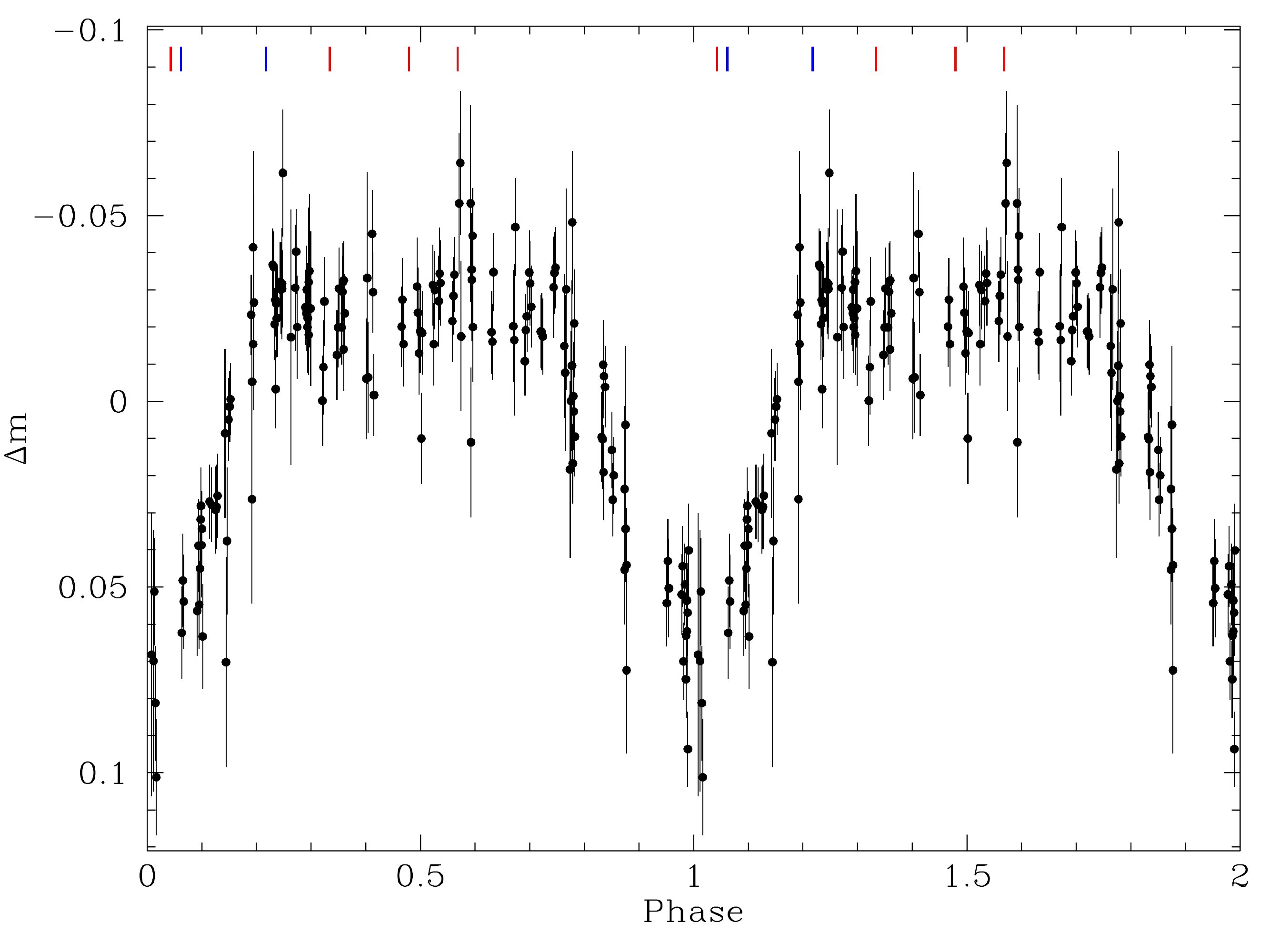}
\caption{Amplitude spectrum (left panel, computed from the combined ZTF and LT data) and the light curve of SDSS\,J1219+4715 phase-folded on the 15.26-h period (right panel, LT data only). The red and blue tickmarks above the folded light curve indicate the phases of the \Ha\ and \Hb\ spectra obtained with the GTC, respectively. The amplitude spectrum (black) shows a complex pattern of aliases, reflecting the window function of the combined ZTF and LT observations. Shown in light blue are the data pre-whitened with a sine wave of $P=15.26$\,h. No signal exceeding the significance threshold (blue dashed line) is detected after the window function is removed.}
\label{f:photometry}
\end{figure*}

\subsection{Photometric and spectroscopic variability} 
\label{s:variability}
The combined LT and ZTF light curve of SDSS\,J1219+4715 extends from 2018 March 25 to 2020 July 6, with a total of 437 photometric epochs. We computed a discrete Fourier transform of the photometry using the \textsc{tsa} context within \textsc{midas}. The resulting amplitude spectrum (Fig.\,\ref{f:photometry}, left panel) shows a number of sharp signals spaced out by one-day aliases that are typical of single-site observations. The strongest alias corresponds to a period of $P=15.26415\pm0.00019$\,h, where the uncertainty was determined from a sine fit to the data. We subjected the photometric data to a bootstrap test \citep[see][]{southworthetal06-1, southworthetal07-2} and found a 99.9~per cent likelihood that the strongest alias correctly identifies the underlying period. Pre-whitening the light curve with that period and computing a new amplitude spectrum completely removes all significant signals (Fig.\,\ref{f:photometry}, left panel).

The limits on the mass and luminosity of a companion to SDSS\,J1219+4715 (Section\,\ref{s:companion}) rule out that the photometric modulation is associated with binarity, and the most likely interpretation is that it reflects the spin period of the white dwarf. 

The phase-folded light curve of SDSS\,J1219+4715 (Fig.\,\ref{f:photometry}, right panel) shows a symmetric parabola-shaped minimum that extends in phase, $\Delta\phi\simeq0.4$. Whereas the ZTF $g$- and $r$-band data, because of their long baseline, are useful in improving the accuracy of the spin period, their photometric precision is insufficient to assess any colour dependence of the modulation. 

\begin{figure}
\centering{\includegraphics[width=\columnwidth]{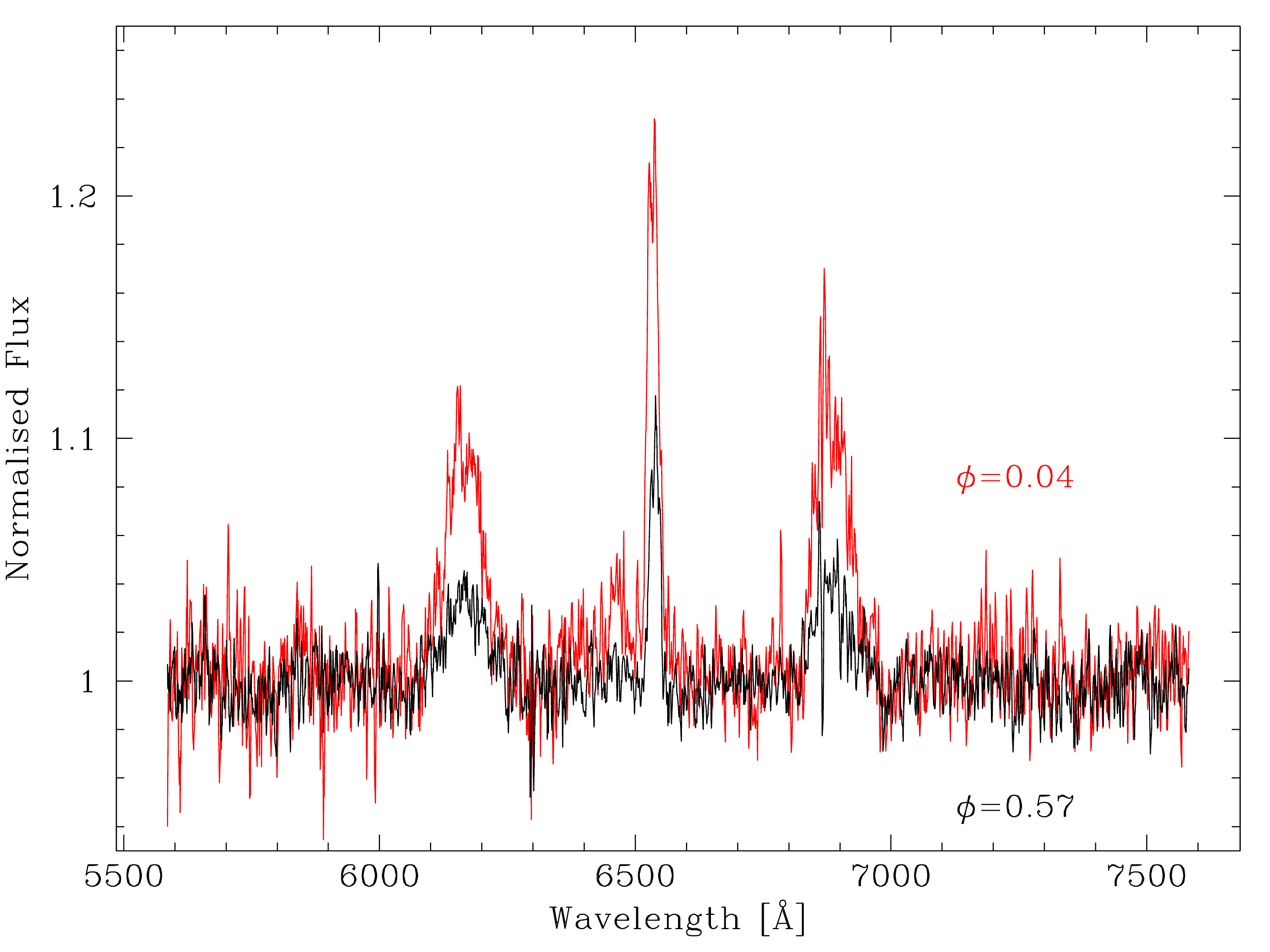}}
\caption{Normalised GTC \Ha\ spectra of SDSS\,J1219+4715 obtained close to the photometric minimum (red) and maximum (black). The strength of the emission lines varies in anti-phase with the broad-band photometry, however, the Zeeman components do not shift in wavelength.}
\label{f:ha_stack}
\end{figure}

We established a photometric ephemeris for SDSS\,J1219+4715 where phase zero corresponds to the minimum brightness of the system, 
\begin{equation}
T_0(\mathrm{HJD}) = 2458202.857(1) + 0.6360061(77) \times E~.
\end{equation}
We determined the zero-point by fitting a parabola to the broad minimum in the phase-folded light curve. Using this ephemeris, we computed the phases of the GTC spectra (see Fig.\,\ref{f:photometry}, left panel). While our spectroscopic sampling of the spin phase is limited, the data suggest that the strength of the emission lines varies in anti-phase with the photometry, i.e. the \Ha\ emission is strongest during the photometric minimum (Fig.\,\ref{f:ha_stack}). We integrated the \Ha\ emission line fluxes after subtracting the underlying continuum using a polynomial fit, and find that the strength of the \Ha\ emission line varies from $(5.2\pm0.1)\times10^{-15}~\mathrm{erg\,cm^{-2}\,s^{-1}}$ near the photometric minimum ($\phi=0.04$) to $(1.2\pm0.1)\times10^{-15}~\mathrm{erg\,cm^{-2}\,s^{-1}}$ near the photometric maximum ($\phi=0.57$). Adopting an average \Ha\ flux of $3.2\times10^{-15}~\mathrm{erg\,cm^{-2}\,s^{-1}}$ results in a luminosity of $L(\Ha)\simeq1.9\times10^{27}~\mathrm{erg\,s^{-1}}$, which is almost identical to the $\simeq2\times10^{27}~\mathrm{erg\,s^{-1}}$ reported by  \citet{greenstein+mccarthy85-1} for GD356. Also the emission-line flux of \Hb\ varies in anti-phase with the photometric modulation,  $(2.8\pm0.2)\times10^{-15}~\mathrm{erg\,cm^{-2}\,s^{-1}}$ at $\phi=0.06$ and $(1.6\pm0.0.5)\times10^{-15}~\mathrm{erg\,cm^{-2}\,s^{-1}}$ at $\phi=0.21$ (the latter spectrum being rather noisy as it was taken under poor conditions). We note that the same anti-phased behaviour between the emission lines and the broad-band continuum was observed in SDSS\,J1252$-$0234 \citep{redingetal20-1}~--~with the noticeable difference that in this system the emission lines fade completely during the photometric maximum, revealing the photospheric Balmer absorption lines of the white dwarf.

The fact that the emission-line fluxes are anti-phased with the photometric modulation complicates the interpretation of the latter one, as it is apparently not associated with the additional flux contained in the emission lines. Many single magnetic white dwarfs that do not show Balmer emission lines are known to exhibit spectroscopic and photometric variability as a function of their spin phase \citep{brinkworthetal05-1, euchneretal06-1, brinkworthetal13-1, hermesetal17-2}, related to subtle changes in the photospheric spectrum due to the changing viewing geometry of the magnetic field distribution. Accurate, simultaneous spectroscopic (ideally spectropolarimetric) and photometric observations will be required to disentangle the variability intrinsic to the white dwarf and the associated with the emitting region. 

\begin{table*}
\newcommand{\mc}[3]{\multicolumn{#1}{#2}{#3}}
\caption{\label{t:3mwd} Stellar parameters of the three magnetic white dwarfs exhibiting Zeeman-split Balmer emission lines.}
\centering
\begin{tabular}{llr@{$\,\pm\,$}lr@{$\,\pm\,$}lr@{$\,\pm\,$}lr}\hline
\mc{2}{l}{Parameter} & 
\mc{2}{c}{GD356} & 
\mc{2}{c}{SDSS\,J1252--0234} & 
\mc{2}{c}{SDSS\,J1219+4715} &
References \\
\hline
Parallax               & $\varpi$\,[mas]         & 49.65     & 0.03  & 12.94    & 0.11   & 14.28   & 0.12   & 1\\
Distance               & $d$\,[pc]               & 20.1      & 0.1   & 77.1     & 0.7    & 69.6    & 0.6    & 2\\
\textit{Gaia} photometry & $G$\,[mag]            & 14.9808	 & 0.0007 & 17.4775	& 0.0019 & 17.5612 & 0.0018 & 1\\
Proper motion          & $\mu_\alpha$\,[mas/yr]  & $-$119.40 & 0.06  & 49.64    & 0.25 & $-$113.28 & 0.11   & 1\\	
                       & $\mu_\delta$\,[mas/yr]  & $-$190.61 & 0.08	 & $-$39.95 & 0.16 & $-$3.492  & 0.15   & 1\\
Effective temperature  & \Teff\,[K]              & 7698      & 74    & 7856     & 101  & 7500      & 148    & 3\\
Surface gravity        & \logg\ (cgs)            & 8.22      & 0.04  & 7.98     & 0.06 & 8.09      & 0.04   & 3\\
Mass                   & \Mwd\,[\Msun]           & 0.733     & 0.023 & 0.583    & 0.031& 0.649     & 0.022  & 3\\ 
Cooling age            & $\tau_\mathrm{cool}$\,[Myr] & 1916  & 144   & 1136     & 95   & 1558      & 124    & 3\\  
Magnetic field$^*$     & $B$\,[MG]               & 11        & 1.1   & 5        & 0.1  & 18.5      & 1.0    & 4,5,3 \\
Spin period            & $P$\,[h]                & 1.9280 & 0.0011   & 0.0881328 & 0.0000036  & 15.26415 & 0.00019 & 6,5,3 \\
\hline
\end{tabular}
\begin{minipage}{\textwidth}
$^*$~We report the measurements obtained from the emission lines, which correspond to the field strength in the emitting region. 
$^1$~\citet{gaiaetal18-1}; 
$^2$~\citet{bailer-jonesetal18-1};
$^3$~this paper;
$^4$~\citet{greenstein+mccarthy85-1};
$^5$~\citet{redingetal20-1};
$^6$~\citet{brinkworthetal04-1}, we report the average and standard deviation of their two possible periods.
\end{minipage}
\end{table*}

\section{Discussion}

\subsection{A closely clustered class of white dwarfs}
\label{s:clustered}
\begin{figure}
\centering{\includegraphics[width=\columnwidth]{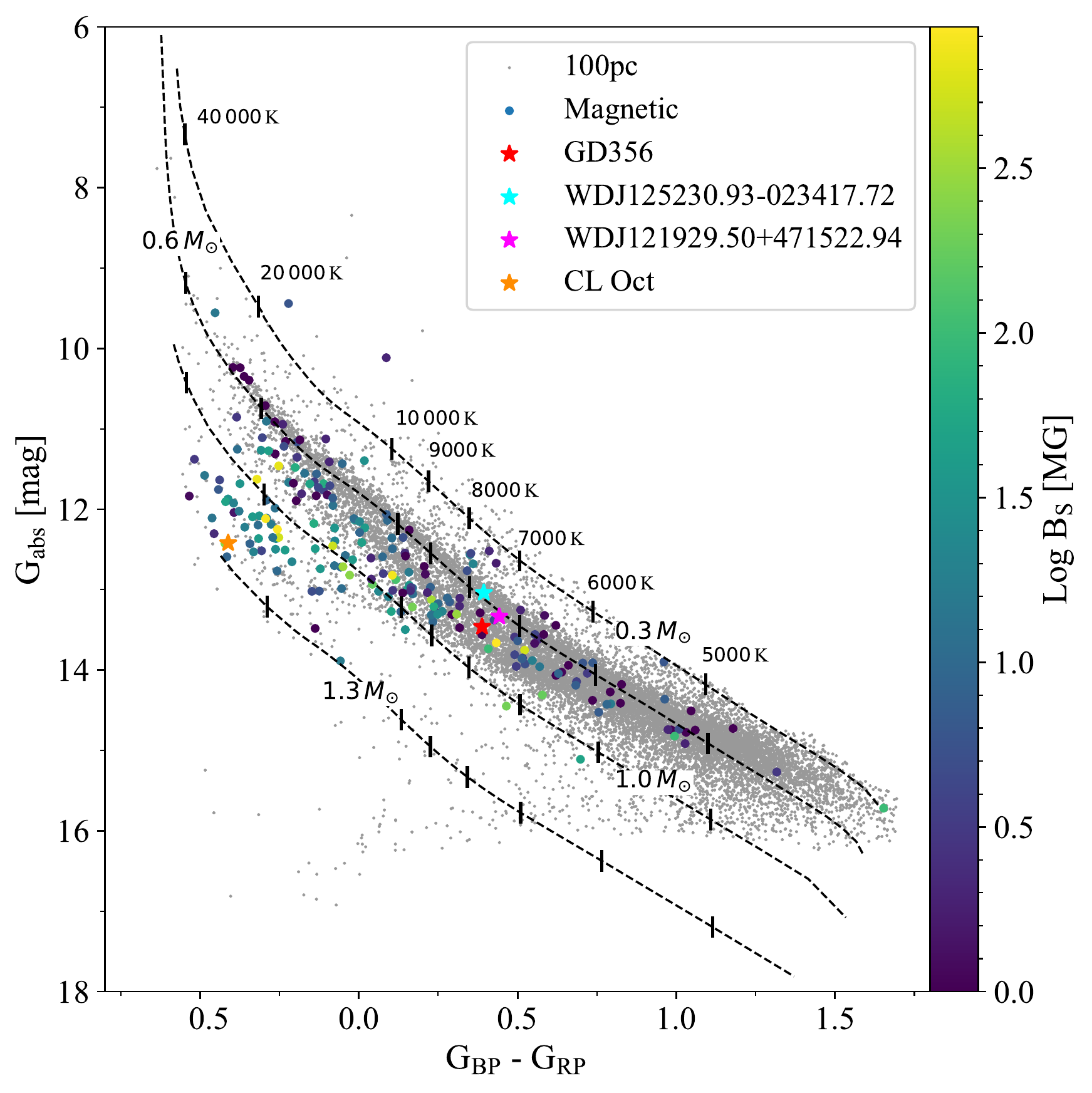}}
\caption{Hertzsprung-Russell diagram of the white dwarfs within 100\,pc \citep{gentile-fusilloetal19-1}, with those known to be magnetic \citep{ferrarioetal15-1, hollandsetal17-1, mccleeryetal20-1} colour-coded by their field strength. The three magnetic white dwarfs exhibiting Zeeman-split emission lines cluster extremely closely within the parameter space spanned by the overall population of magnetic white dwarfs. White dwarf cooling tracks$^{\ref{fn:bergeron}}$ for a range of white dwarf masses are superimposed, with temperatures along those tracks indicated as short vertical tick marks. For comparison, the massive ($M_\mathrm{wd}\simeq1.3$\,\Msun), rapidly spinning ($P=725$\,s) and strongly magnetic ($B\simeq100-800$\,MG) white dwarf CL\,Oct which is discussed as a merger product (\citealt{barstowetal95-2, ferrarioetal97-1, kuelebietal10-1}), is shown as an orange star~--~clearly differing from the three stars discussed here.}
\label{f:hrd}
\end{figure}

Inspection of the physical characteristics of GD356, SDSS\,J1252$-$0234, and SDSS\,J1219+4715 (Table\,\ref{t:3mwd}) and their location in the \textit{Gaia} Hertzsprung-Russell diagram (Fig.\,\ref{f:hrd}) shows they cluster closely in effective temperature ($\simeq7500-7860$\,K), mass ($\simeq0.58-0.73$\,\Msun), cooling age ($\simeq1.1-1.9$\,Gyr), and magnetic field strength ($\simeq5-19$\,MG) when compared to the full parameter space occupied by white dwarfs. The tangential velocities of the three stars are $\simeq20-40$~km\,s$^{-1}$, well within the range of white dwarfs with similar cooling ages \citep[e.g.][]{mccleeryetal20-1}, consistent with thin disc membership, and not indicative of any major dynamical interactions in their past lives. A noticeable exception are their spin periods, which span over two orders of magnitude. Non-magnetic single white dwarfs have typical rotation periods of several tens of hours, with very few stars known to spin faster than $\simeq5$\,h \citep{hermesetal17-1}. SDSS\,J1252$-$0234 is clearly an extreme outlier, requiring an explanation for its very short rotation period (see Section\,\ref{s:planetaccretion}).

This clustering begs the question: is it representative of an observational selection effect, or reflective of the physical mechanism underlying the still unknown process causing the emission lines? We discuss in Section\,\ref{s:emissionlineorigin} that the spot luminosity is unlikely to be related to ongoing accretion, and we already noticed (Section\,\ref{s:variability}) that the strength of the Balmer emission-line fluxes in these three stars is similar.

Assuming that these line strengths are typical for the mechanism that produces them, the emission lines would be harder to be detected against the photospheres of hotter, brighter white dwarfs. To simulate this effect, we extracted the \Ha\ line fluxes from the GTC average spectrum by subtracting the continuum via a polynomial fit. We then produced a set of DA white dwarf models with $\log g=8.09$ and temperatures ranging from 6000\,K to 10\,000\,K, scaled them to a distance of 69.6\,pc, and added Gaussian noise to emulate the data quality of the GTC spectra. Finally, we added the observed emission-line fluxes to the (noisy) white dwarf models, and subjected them both to a visual inspection and an equivalent width measurement of the \Ha\ region. The conclusion from this exercise was that the \Ha\ line fluxes seen in SDSS\,J1219+4715 would be detectable in white dwarfs with temperatures of up to $\simeq9000$\,K~--~corresponding to a cooling age of $\simeq1$\,Gyr, or $\simeq60$\% of the present cooling age of SDSS\,J1219+4715. Assuming that the mechanism that generates the emission lines is always active, there is hence a reasonably long period of time where they could be detected at slightly hotter effective temperatures, and a similar reasoning holds true for the other two systems. It therefore appears somewhat coincidental that all three white dwarfs are found at ages which are clearly past the detection threshold~--~and there is no shortage of slightly hotter and younger magnetic white dwarfs (Fig.\,\ref{f:hrd}). 

However, the question of observational selection effects becomes more problematic at lower temperatures~--~as the white dwarf luminosity decreases, it should be \textit{easier} to detect emission lines of similar strength among cooler magnetic white dwarfs. There is a substantial number of magnetic white dwarfs with $\Teff\lesssim7000$\,K, and emission lines were detected in none of those \citep{ferrarioetal15-1, hollandsetal15-1, hollandsetal17-1, landstreet+bagnulo19-1, mccleeryetal20-1, kawka20-1}. It appears hence that the mechanism causing the emission lines in the three stars discussed here is no longer active among cooler magnetic white dwarfs.

Thus, taking the clustering of these three stars at face value suggests that whatever the mechanism responsible for the emission lines is, it becomes active at $\Teff\simeq7700$\,K, and is relatively short-lived, $\simeq0.5-1$\,Gyr. A quick comparison of the rotational kinetic energy of the white dwarfs with the estimated luminosity of the emission lines rules out that the emission process itself results in a sufficiently rapid spin-down of the white dwarf that might eventually stop the mechanism. Within the unipolar inductor model involving a planet in a close-in orbit, the limited life time of the planet due to ohmic dissipation may provide a natural explanation for this short duration \citep{lietal98-1, veras+wolszczan19-1}. However, the late appearance of this effect along the white dwarf cooling track still remains a mystery.

\subsection{Origin of the emission lines}
\label{s:emissionlineorigin}
The detection of photospheric emission lines from \textit{single} white dwarfs is extremely rare, with the exception of extremely hot and young (pre-) white dwarfs \citep[e.g.][]{werner91-1, werneretal91-1} and a number of helium-atmosphere white dwarfs with $\Teff\simeq14\,000-17\,000$\,K \citep{kleinetal20-1}~--~in both cases, the presence of the emission lines is intrinsic to their atmospheric structures. 

To our knowledge, only the following seven cooler white dwarfs display noticeable emission lines: the three magnetic white dwarfs discussed here (optical Balmer lines, Table\,\ref{t:3mwd}), one nearby white dwarf with weak, single-peaked \Ha\ emission \citep{tremblayetal20-1, mccleeryetal20-1}\footnote{The sharp \Ha\ emission in WD\,J041246.85+754942.26 showed no radial velocity variation in observations obtained on three consecutive nights, and the optical-to-infrared spectral energy distribution rules out a companion earlier than a T-type brown dwarf. At the moment, the origin of this emission line remains unexplained, and the association with the stars discussed here unclear. With $\Teff\simeq8500$\,K, $\log g\simeq8.25$, WD\,J0412+7549 lies close to the three magnetic white dwarfs exhibiting Balmer line emission, and one speculative hypothesis is that it might have a weak field ($B\lesssim100$\,kG). Additional observations of this white dwarf are encouraged.}, one white dwarf accreting planetary debris (\Ion{Ca}{ii}~H/K lines, PG\,1225$-$075, \citealt{kleinetal10-1}), and two white dwarfs with carbon-dominated (DQ) atmospheres (metal lines in the far-ultraviolet, \citealt{provencaletal05-1}).

\citet{greenstein+mccarthy85-1} speculated about either chromospheric activity resulting from interactions between convective instabilities and the magnetic field, or accretion causing the emission lines in GD356. Given that the strong magnetic field in GD356 is suppressing convection \citep{tremblayetal15-1, gentile-fusilloetal18-1}, chromospheric activity is unlikely. 

Sharp photospheric emission lines are indeed detected among a number of close binaries containing an accreting white dwarf. Examples include several AM\,CVn systems, ultra-short period  binaries consisting of a cool white dwarf accreting from an extremely low-mass degenerate donor \citep{marsh99-1, kupferetal16-1}, and detached white dwarf plus M-dwarf binaries, in which the white dwarf accretes from the stellar wind of its companion \citep{tappertetal07-1, tappertetal11-2}. One characteristic that all these systems have in common is that they harbour ($\lesssim10\,000$\,K) cool white dwarfs that accrete at relatively low rates, and whereas no quantitative model has been developed, it appears likely that it is the energy deposited by the accreted material that results in a temperature inversion in the white dwarf atmosphere, resulting in the observed emission lines.  

The system that bears most resemblance to the three magnetic white dwarfs discussed here is SDSS\,J030308.35+005444.1, a short-period detached binary containing a ($\lesssim8000$\,K) cool  white dwarf plus an M-dwarf \citep{pyrzasetal09-1}. \citet{parsonsetal13-1} detected Zeeman-split Balmer emission lines, derived a field strength of $\simeq8$\,MG, and argued that the emission lines are caused by accretion of wind captured by the white dwarf and funnelled towards its magnetic pole(s). A key feature of accretion onto magnetic white dwarfs is that the shock-heated plasma near the white dwarf surface also emits cyclotron radiation \citep{visvanathan+wickramasinghe79-1, woelk+beuermann92-1}. At field strengths of $\sim10$\,MG, the cyclotron emission lines are located in the near-infrared, and  are indeed detected in SDSS\,J030308.35+005444.1 as an infrared excess over the M-dwarf \citep{debesetal12-3}.

Given their field strengths of $5-19$\,MG (Table\,\ref{t:3mwd}), accretion onto the three single magnetic white dwarfs with Zeeman-split Balmer emission lines should equally result in cyclotron emission, with an associated infrared excess~--~which is, however, detected in none of them. This strongly argues against ongoing accretion as the cause for the observed emission lines. This extends the conclusions of \citet{weisskopfetal07-1}, who placed an upper limit on the X-ray luminosity of GD356 of $L_\mathrm{x}\le6\times10^{25}\,\mathrm{erg\,s^{-1}}$, and already argued that the non-detection of an infrared excess at GD356 is inconsistent with the possible presence of a hot corona as a source of the Balmer emission lines (see also \citealt{musielaketal95-1} for an earlier X-ray study of GD356 and \citealt{ferrarioetal97-2} for an additional discussion arguing against ongoing accretion). No X-ray observations of the other two stars discussed here are available. 

With both chromospheric activity and accretion being unlikely to be the cause for the Balmer emission lines in these three magnetic white dwarfs, the unipolar inductor model \citep{lietal98-1, wickramasingheetal10-1} remains currently the only plausible explanation. This model requires a conductive planet or planet core in a close orbit around each of the three stars, which, in the light of the detections of solid planetesimals \citep{vanderburgetal15-1, manseretal19-1, vanderboschetal20-1} as well as giant planets \citep{gaensickeetal19-1}, appears entirely feasible. The fact that all three stars discussed here are relatively nearby ($d\lesssim80$\,pc) implies that such configurations of evolved planetary systems may not be exceedingly rare. 

\subsection{A double-degenerate merger origin?}
\label{s:merger}
\citet{wickramasingheetal10-1} and \citet{redingetal20-1} argued that GD356 and SDSS\,J1252$-$0234 could originate from double-degenerate mergers, which would explain both the magnetic field and the short rotation periods of their white dwarfs. In addition, second-generation planets may form out of the metal-rich debris disc left behind by the merger event \citep{garcia-berroetal07-1}, providing the key component for the unipolar inductor model of \citet{lietal98-1}. 

The masses of the three magnetic white dwarfs showing Balmer emission lines, even when accounting for the fact that we may slightly underestimate them (Sect.\,\ref{s:wdparam}), are close to the average mass of the field white dwarf population\footnote{The average white dwarf mass in a magnitude-limited sample is slightly lower, $<M_\mathrm{wd}>\simeq0.6$\,\Msun, as lower mass white dwarfs are larger and hence can be detected out to larger distances \citep{koesteretal79-1, bergeronetal92-1, bergeronetal19-1, tremblayetal19-2}.}, $<M_\mathrm{wd}>\simeq0.65$\,\Msun\ \citep{giammicheleetal12-1, hollandsetal18-2, tremblayetal20-1}, and hence if produced by double-degenerate mergers, the progenitor binaries would have contained two He-core white dwarfs. Detailed hydro-dynamical simulations of double-degenerate mergers show that He-core mergers eject very little mass ($\lesssim2\times10^{-3}$\,\Msun). This implies that the total masses of the double-degenerates that merged were very close to the masses measured for the three magnetic white dwarfs, i.e. $\simeq0.58-0.73$\,\Msun. 

Inspecting the roster of double-degenerates with measured masses for both components \citep{rebassa-mansergasetal17-1, breedtetal17-1, napiwotzkietal20-1} suggests that potential progenitor systems matching the top-end of the mass range spanned by the three magnetic systems are known, but that they are rare~--~and their merger time scales are typically exceeding a Hubble time \citep{geieretal10-1}. Given the large observational bias favouring the identification of lower mass white dwarfs (because of their larger radii), the scarcity of progenitors is exacerbated. We note that progenitors containing one extremely low mass (ELM) white dwarf are even less likely, as the total masses of the known ELM binaries peak\footnote{Exceptions exist, but are \textit{extremely} rare \citep{brownetal20-1}.} at $\simeq1$\,\Msun\ \citep{brownetal16-1}, and as ELM binaries are intrinsically rare \citep{kawkaetal20-1}.

One additional potential issue with invoking second-generation planets forming out of the ejected material is that \citet{garcia-berroetal07-1} modelled a CO+He merger. It is not clear if a He+He merger would result in sufficiently metal-rich ejecta to form a conductive, second-generation planet. 

Assuming nevertheless a merger origin for these three magnetic white dwarfs, the characteristics of the known population of double-degenerates would suggest that a larger number of similar systems with resulting masses for the merger product of $\simeq0.8-1.0$\,\Msun\ should exist~--~yet none has been found so far. Moreover, it would appear very coincidental that these three stars share many of their properties, in particular their effective temperatures and cooling ages, as discussed in the previous section. A double-degenerate merger will result in a re-heated and possibly strongly magnetic white dwarf \citep{garcia-berroetal12-1, wegg+phinney12-1}, such as it may be the case for CL\,Oct (see Fig.\,\ref{f:hrd}, \citealt{ferrarioetal97-1}). If the Balmer line mechanism in such a system is active, it should be detectable before it cools down to $\simeq7700$\,K (Section\,\ref{s:clustered} and Fig.\,\ref{f:hrd}).

We conclude that while a double-degenerate merger origin cannot be ruled out, the clustered properties of these three magnetic white dwarfs and the sufficiently disparate characteristics of the known double-degenerates argue against it. One final note concerns the cooling ages calculated in Sect.\,\ref{s:wdparam}. In the case of a double-degenerate evolution history, these would approximate the time since the merger event \citep{wegg+phinney12-1, temminketal19-1}, but that would change none of our arguments regarding the tight clustering of the three stars in their physical properties outlined above and in Sect.\,\ref{s:clustered}.

\subsection{White dwarf spin-up via accretion from a giant planet?}
\label{s:planetaccretion}
Our argument against double-degenerate mergers leaves us with the question: how can some single white dwarfs reach very short periods, such as SDSS\,J1252$-$0234? A possible answer is, again, linked to planets surviving and scattering into close-in orbits around the white dwarf. \citet{stephanetal20-1} demonstrated that ingestion of a giant planet can spin-up white dwarfs to close to break-up, thus reaching spin periods in the range of minutes appears entirely possible. The discovery of a giant planet at WD\,J091405.30+191412.25 \citep{gaensickeetal19-1} corroborates that this is a plausible scenario~--~moreover, if only the gaseous envelope of the giant planet is accreted, but its (conductive) core survives, it naturally provides the key ingredient for the unipolar inductor model.

\section{Conclusions}
We have identified SDSS\,J1219+4715 as an additional magnetic white dwarf exhibiting Zeeman-split Balmer emission lines and photometric variability with a 15.26-h period, which we interpret as being related to the white dwarf rotation. The upper limit on the temperature of a possible substellar or giant planet companion is $\simeq250$\,K.  The emission lines originate in a region with a fairly homogeneous field strength of $B\simeq18.5\pm1.0$\,MG. Whereas the emission lines vary in flux over the  white dwarf spin phase by a factor approximately four, they always remain visible, indicating that the emitting region never fully disappears behind the limb of the white dwarf. With a temperature of $7500\pm148$\,K, a mass of $0.649\pm0.022$\,\Msun, and a cooling age of $1.5\pm0.1$\,Gyr, SDSS\,J1219+4715 very closely resembles the other two members of this small class of stars: GD356 and SDSS\,J1252$-$0234. We argue that this clustering in physical parameters, including the strength of the emission-line fluxes, suggests a common mechanism that (a) becomes active at cooling ages of $\simeq1.5$\,Gyr, (b) lasts $\simeq0.5-1.0$\,Gyr, and (c) does not affect all magnetic white dwarfs. Given the growing observational evidence for the existence of planetesimals and planets in close orbits around white dwarfs, the unipolar inductor model developed for GD356 seems a plausible scenario that satisfies (b) and (c) above, however, the onset of the emission lines at advanced cooling ages remains unexplained. We encourage closer spectroscopic scrutiny of the known magnetic white dwarfs with temperatures $6000-9000$\,K to either detect additional examples of this class of stars, or to place stringent upper limits on the presence of Balmer emission lines.

\section*{Acknowledgements}
We thank Detlev Koester and Ken Shen for insightful discussions, and the referee S.O. Kepler for helpful comments. BTG was supported by the UK STFC grant ST/T000406/1. {PR-G} acknowledges support from the State Research Agency (AEI) of the Spanish Ministry of Science, Innovation and Universities (MCIU), and the European Regional Development Fund (FEDER) under grant AYA2017--83383--P.  The research leading to these results has received funding from the European Research Council under the European Union's Horizon 2020 research and innovation programme n. 677706 (WD3D). The use of the {\sc pamela} and {\sc molly} packages developed by Tom Marsh is acknowledged.

Based on observations made with the Gran Telescopio Canarias (GTC) installed in the Spanish Observatorio del Roque de los Muchachos of the Instituto de Astrof\'isica de Canarias, in the island of La Palma, and on observations made with the Liverpool Telescope, which is operated on the island of La Palma by Liverpool John Moores University in the Spanish Observatorio del Roque de los Muchachos of the Instituto de Astrof\'isica de Canarias with financial support from the UK Science and Technology Facilities Council. 

Funding for the Sloan Digital Sky Survey IV has been provided by the Alfred P. Sloan Foundation, the U.S. Department of Energy Office of Science, and the Participating Institutions. SDSS-IV acknowledges support and resources from the Center for High-Performance Computing at the University of Utah. The SDSS web site is www.sdss.org.

SDSS-IV is managed by the Astrophysical Research Consortium for the Participating Institutions of the SDSS Collaboration including the Brazilian Participation Group, the Carnegie Institution for Science, Carnegie Mellon University, the Chilean Participation Group, the French Participation Group, Harvard-Smithsonian Center for Astrophysics,  Instituto de Astrof\'isica de Canarias, The Johns Hopkins University, Kavli Institute for the Physics and Mathematics of the Universe (IPMU) /  University of Tokyo, the Korean Participation Group, Lawrence Berkeley National Laboratory, Leibniz Institut f\"ur Astrophysik Potsdam (AIP),  Max-Planck-Institut f\"ur Astronomie (MPIA Heidelberg), Max-Planck-Institut f\"ur Astrophysik (MPA Garching), Max-Planck-Institut f\"ur Extraterrestrische Physik (MPE), National Astronomical Observatories of China, New Mexico State University, New York University, University of Notre Dame, Observat\'orio Nacional / MCTI, The Ohio State University, Pennsylvania State University, Shanghai Astronomical Observatory, United Kingdom Participation Group, Universidad Nacional Aut\'onoma de M\'exico, University of Arizona, University of Colorado Boulder, University of Oxford, University of Portsmouth, University of Utah, University of Virginia, University of Washington, University of Wisconsin, Vanderbilt University, and Yale University.

The Pan-STARRS1 Surveys (PS1) and the PS1 public science archive have been made possible through contributions by the Institute for Astronomy, the University of Hawaii, the Pan-STARRS Project Office, the Max-Planck Society and its participating institutes, the Max Planck Institute for Astronomy, Heidelberg, and the Max Planck Institute for Extraterrestrial Physics, Garching, The Johns Hopkins University, Durham University, the University of Edinburgh, the Queen's University Belfast, the Harvard-Smithsonian Center for Astrophysics, the Las Cumbres Observatory Global Telescope Network Incorporated, the National Central University of Taiwan, the Space Telescope Science Institute, the National Aeronautics and Space Administration under Grant No. NNX08AR22G issued through the Planetary Science Division of the NASA Science Mission Directorate, the National Science Foundation Grant No. AST-1238877, the University of Maryland, Eotvos Lorand University (ELTE), the Los Alamos National Laboratory, and the Gordon and Betty Moore Foundation.

Based on observations obtained with the Samuel Oschin 48-inch Telescope at the Palomar Observatory as part of the Zwicky Transient Facility project. ZTF is supported by the National Science Foundation under Grant No. AST-1440341 and a collaboration including Caltech, IPAC, the Weizmann Institute for Science, the Oskar Klein Center at Stockholm University, the University of Maryland, the University of Washington, Deutsches Elektronen-Synchrotron and Humboldt University, Los Alamos National Laboratories, the TANGO Consortium of Taiwan, the University of Wisconsin at Milwaukee, and Lawrence Berkeley National Laboratories. Operations are conducted by COO, IPAC, and UW.

This publication makes use of data products from the Wide-field Infrared Survey Explorer, which is a joint project of the University of California, Los Angeles, and the Jet Propulsion Laboratory/California Institute of Technology, funded by the National Aeronautics and Space Administration.

\section*{Data Availability}
All observational data used in this paper will be available from the public data archives of the SDSS, LT, GTC, and ZTF.





\bibliographystyle{mnras}
\bibliography{aamnem99,aabib}







\bsp	
\label{lastpage}
\end{document}